\documentclass[12pt,preprint]{aastex}
\usepackage{gensymb} 
\usepackage[version=4]{mhchem}
\usepackage{amsmath}
\usepackage{adjustbox}
\usepackage[colorlinks=true,citecolor=blue,breaklinks]{hyperref}
\shorttitle{\ce{H2O} in the atmosphere of Jupiter}
\shortauthors{Manna \& Pal}

\begin{document}

\title{ALMA spectroscopic detection of water vapour in the atmosphere of the giant gas planet Jupiter}

\author{Arijit Manna}

\affil{Midnapore City College, Kuturia, Bhadutala, Paschim Medinipur, West Bengal, India 721129} 

\and

\author{Sabyasachi Pal}

\affil{Indian Centre for Space Physics, 43 Chalantika, Garia Station Road, Kolkata, India 700084\\
Midnapore City College, Kuturia, Bhadutala, Paschim Medinipur, West Bengal, India 721129}

\begin{abstract}
In the Jovian atmosphere, the trace species are detected for the first time after the collision of comet Shoemaker-Levy 9 near 44$^\circ$S in July 1994. Most of these trace species are detected with spectroscopic millimeter/submillimeter observation. In the atmosphere of Jupiter, trace gases play an important role in atmospheric chemistry with heterogeneous and homogeneous chemical reactions, interaction with radiation, and phase transition. Here we present the first spectroscopic detection of the rotational emission line of water (\ce{H2O}) in the atmosphere of Jupiter at frequency $\nu$ = 183.310 GHz with molecular transition J = 3$_{1,3}$--2$_{2,2}$ using Atacama Large Millimeter/Submillimeter Array (ALMA). The statistical column density of water emission line is N(\ce{H2O})$\sim$4$\times$10$^{15}$ cm$^{-2}$. The rotational emission line of \ce{H2O} is found in the stratosphere of Jupiter with $\geq$3$\sigma$ statistical significance. The column density of \ce{H2O} corresponds to the fractional abundance relative to \ce{H2} is f(\ce{H2O})$\sim$ 4$\times$10$^{-9}$.

\end{abstract}


	\section{Introduction}
\label{sec:intro} 
In 1994, the fragments of comet Shoemaker-Levy 9 collided with the stratosphere of Jupiter (known as SL9 event) and formed many volatile gases like CS, HCN, CO, \ce{H2S}, \ce{CS2}, \ce{S2}, and OCS \citep{Nol95, Lel95, Mor03}. The Jovian atmosphere consists of 71\% hydrogen, 24\% helium, and 5\% other dense elements \citep{Gau81}. The atmosphere of Jupiter also contains a large number of trace gases like methane (\ce{CH4}), ammonia (\ce{NH3}), carbon monoxide (CO), ethane (\ce{C2H6}), hydrogen sulfide (\ce{H2S}), phosphine (\ce{PH3}), sulfur (S), and silicon-based (Si) compounds \citep{Kun04, Fle09}. The outer layer of the atmosphere contains frozen ammonia in crystal form \citep{Mas80}. In the Jupiter atmosphere, a big amount of benzene and other hydrocarbons are found using infrared and ultraviolet measurements \citep{Fri02}. Jupiter is permanently covered with a composed cloud of ammonia crystals and possibly ammonium hydrosulfide \citep{Loe17}. The atmospheric cloud layer is $\sim$50 km deep and consists of two decks of clouds with a thick lower deck and a thin cloudy region.

Infrared spectra of the Short-Wavelength Spectrometer (SWS) of ISO detected the water vapour (between 39.48--44.19 $\mu$m) in the upper atmosphere of four giant planets and Saturn moon Titan in order of (1--30)$\times$10$^{14}$ molecules cm$^{-2}$ which provided the evidence with an external source of oxygen \citep{Feu99}. In the stratosphere of Jupiter, ISO and SWS detected the water vapour with the disk-averaged column density of (2.0$\pm$0.5)$\times$10$^{15}$ cm$^{-2}$ \citep{Lel02}. In the atmosphere of Jupiter, the Juno microwave radiometer detected the abundance of water vapour (with J = 5--6 transitions) at 1.25--22 GHz with approximate pressure 0.7 to 30 bar \citep{Li202}. The detection of water vapour using Juno spacecraft microwave instrument indicated that Jupiter is enhanced in oxygen by roughly three times the solar abundance at the equator which is very important to understand the formation of Jupiter cloud and weather \citep{Bjo20}.

In this paper, we present the spectroscopic detection of water vapour with molecular transition J = 3$_{1,3}$--2$_{2,2}$ at $\nu$ = 183.310 GHz in the atmosphere of Jupiter using ALMA with 7m ACA (Atacama Compact Array) data. In Sect.~\ref{sec:obs}, we briefly describe the observations and data reductions. The result of detection of water from the Jovian atmosphere and abundance of water is presented in Sect.~\ref{sec:result}. The discussion of the detection of water is presented in Sect.~\ref{sec:discussion}.
\begin{figure}
	\centering
	\includegraphics[width=0.49\textwidth]{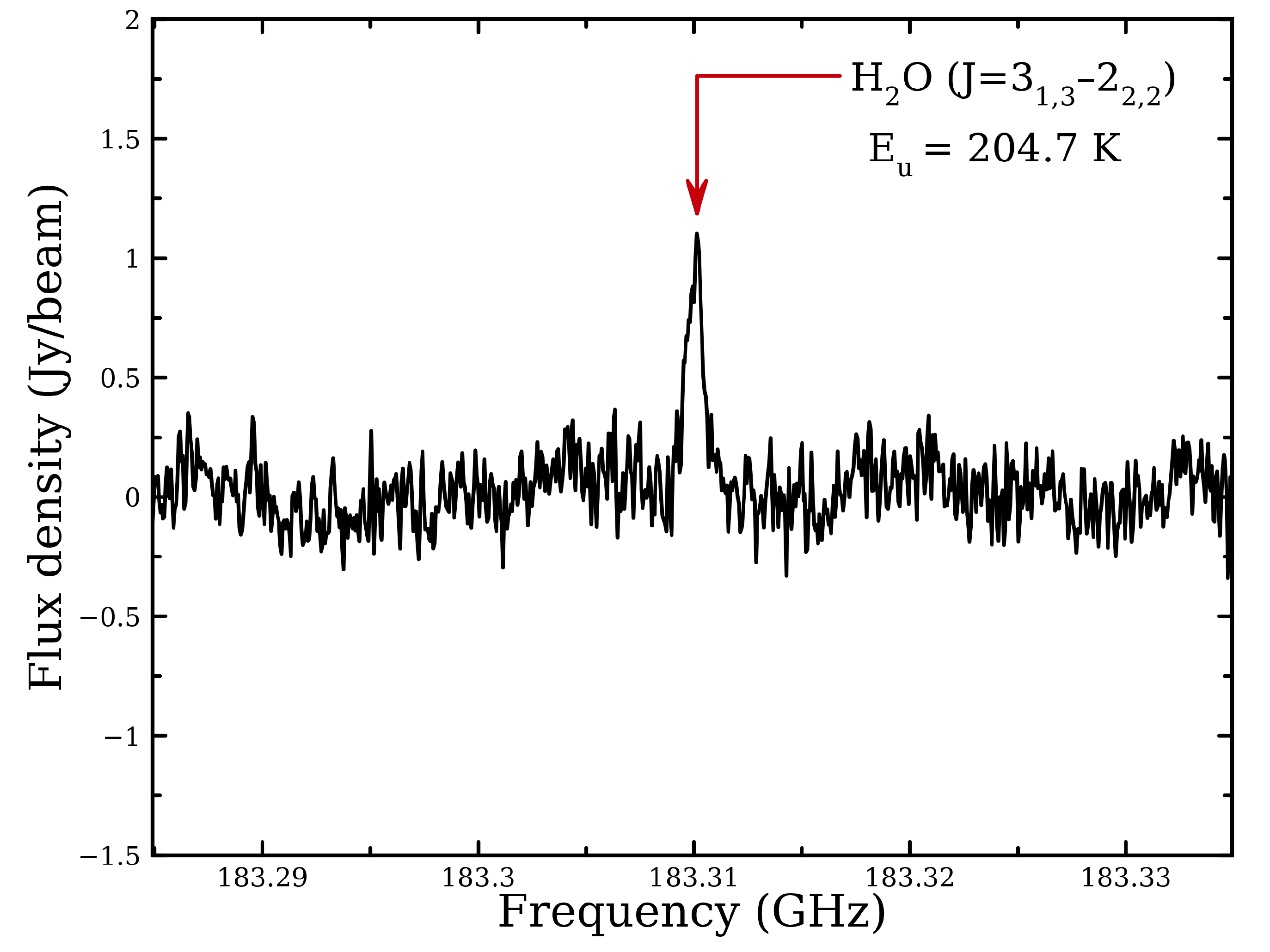}
	\caption{Detection of the emission line of \ce{H2O} at $\nu$ = 183.310 GHz with transition J = 3$_{1,3}$--2$_{2,2}$ on Jupiter at 2018-08-27 with ALMA ACA data. In this spectrum, both vertical and horizontal polarizations are fully averaged. The continuum emission of Jupiter has been subtracted. The spectrum was created for water by ALMA data cubes from the centre of Jupiter (139,822 km at Jupiter distance).}
	\label{fig:water} 
\end{figure}
\section{Observations and data reduction}
\label{sec:obs} 
The solar planet Jupiter was observed with ALMA\footnote{\href{https://almascience.nao.ac.jp/asax/}{https://almascience.nao.ac.jp/asax/}} band 5 receiver in cycle 4 with Atacama Compact Array (7m diameter with short baseline). The observation of the \ce{H2O} line using 183.310 GHz rest frequency in the Jovian atmosphere was done on 27 August 2018 with a 5.6016 AU geocentric distance.A group of the STFC Rutherford Appleton Laboratory (UK), Advanced Receiver Development at Chalmers University and Onsala Space Observatory (GARD, Sweden) are designed and developed the dual-polarization Band 5 receivers which cover the frequency range between 163 and 211 GHz.
The data set initially included four spectral windows and \ce{H2O} rest frequency is set in spectral window 16 with spectral resolution 976.56 kHz. A total of 11 antennas were used during observations of Jupiter. During the observation, J1256--0547 was used as a bandpass calibrator and J1507--1652 was used as phase calibrator. The weather conditions were good, with 1.8 mm of PWV. The illumination factor of Jupiter, during the observation, was 91.42 percent which means 91.42 percent of the mapped planet was in dayside at the time of observation. 

We used standard calibration using the Common Astronomy Software Application\footnote{\href{https://casaguides.nrao.edu/}{https://casaguides.nrao.edu/}} (CASA) with ALMA data for initial data reduction of ACA data. For an improved amplitude calibration, we scaled the datasets to a single reference with Jupiter's angular diameter of 34.42$^{\prime\prime}$.
The continuum flux density for each baseline was scaled and matched with Butler-JPL-Horizons 2012 \citep{But12} for Jupiter model flux for ACA data which is accurate to within 5\%. At first, we measure the $T_{sys}$ for initial data reduction and apply the priori flagging to remove the atmospheric effect of Earth. After the flux, bandpass, and phase calibration, we used the task {\tt MSTRNSFORM} to split the target data into another data set. The background continuum-subtraction of the visibility amplitudes was performed using task {\tt UVCONTSUB} in {\tt CASA} and spectral imaging was carried out using {\tt TCLEAN} task with several numbers of iterations with Hogbom algorithm and natural visibility weighting. The self-calibration process was done using tasks {\tt SELFCAL} and {\tt APPLYCAL} for the improvement in the image sensitivity. The spectral image of Jupiter was transformed from equatorial coordinates to (projected) linear distances concerning the center of Jupiter. The coordinate scale of \ce{H2O} spectrum was corrected for Doppler-shift to Jupiter rest frame using NASA JPL Horizons Topocentric radial velocity.
\begin{figure}
	\centering
	\includegraphics[width=0.49\textwidth]{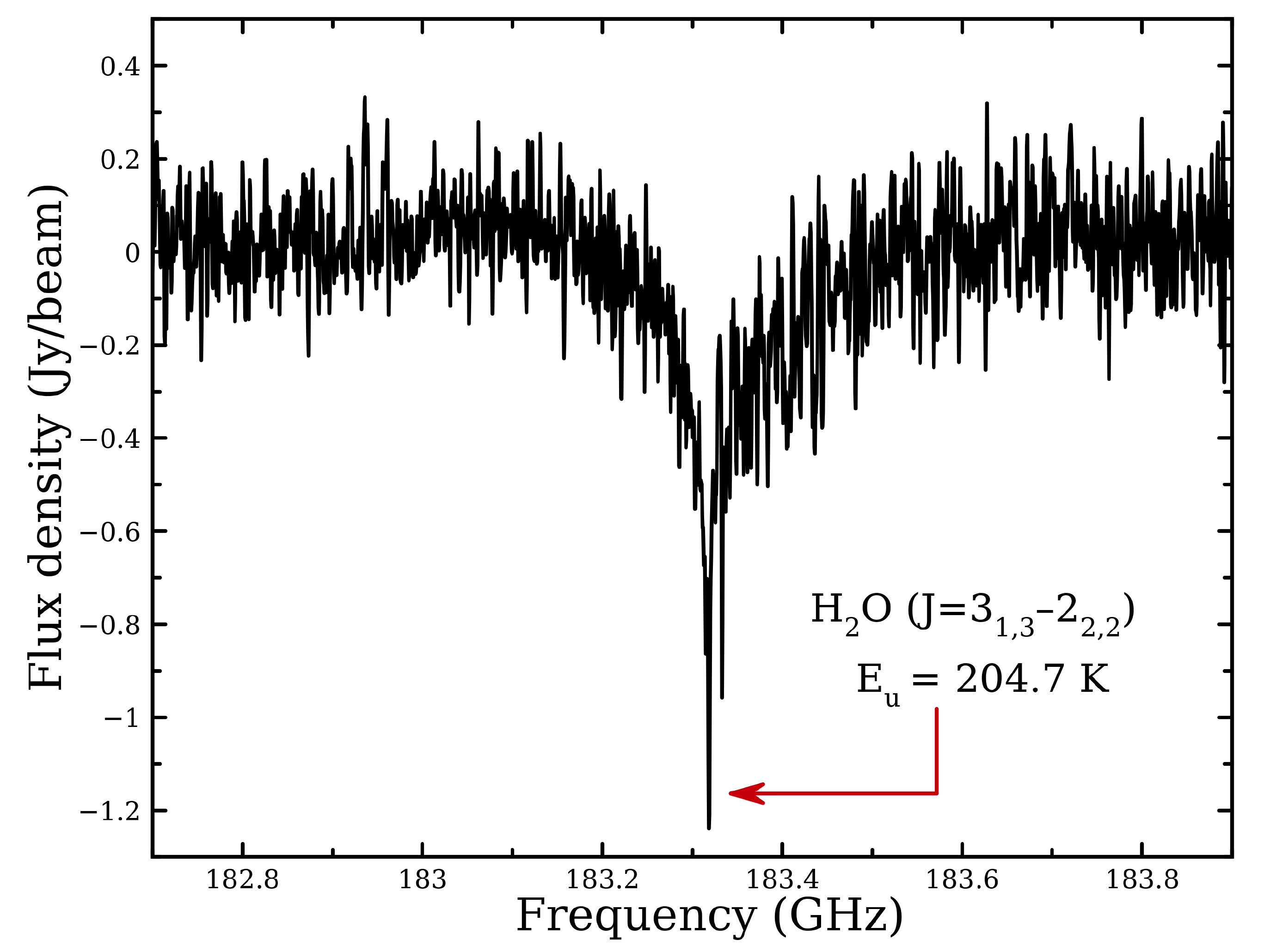}
	\caption{Rotational absorption line of \ce{H2O} at $\nu$ = 183.310 GHz with transition J = 3$_{1,3}$--2$_{2,2}$ find in ALMA Total Power (TP) data of Jupiter. In this spectrum, the strong absorption feature are indicated the earth atmosphere effect on Jupiter single dish data which occour due to strong artifacts in the T$_{sys}$ and target spectra coming from the LO of the PWV radiometers.}
	\label{fig:water1} 
\end{figure}

\section{Result}
\label{sec:result} 
\subsection{Rotational emission line of \ce{H2O} in the atmosphere of Jupiter}
We detect the rotational emission line of water (\ce{H2O}) vapour in the Jovian atmosphere at $\nu$=183.310 GHz with molecular transition J = 3$_{1,3}$--2$_{2,2}$. The rotational emission spectrum of \ce{H2O} in the Jovian atmosphere is shown in Fig.~\ref{fig:water}. The spectral peaks were identified using the online Splatalogue database for astronomical molecular spectroscopy\footnote{\href{https://splatalogue.online//}{https://splatalogue.online//}}. In this spectrum, both vertical and horizontal polarizations are fully averaged. Earlier, the emission lines of water were detected in the stratosphere region of Jupiter using infra-red spectroscopy with ISO and SWAS \citep{Lel02} and sub-mm spectroscopy with Odin space telescope \citep{Be20}. The emission line of water in the Jovian atmosphere using a ground-based telescope is indicated that this water vapour is also probably coming from the stratosphere region. The ALMA detection of water in Jupiter at 183.310 GHz is the first ground-based telescope detection of the molecule. The water is most probably created in the stratosphere of Jupiter due to the collision of comet Shoemaker-Levy 9 in July 1994 near 44$^\circ$S \citep{Nol95, Lel95, Mor03}. 	

\subsection{Earth atmospheric effect in the rotational line of \ce{H2O} in 183 GHz with ALMA Total Power (TP) data}
In the ALMA archive, the other spectroscopic Total Power (TP) data of Jupiter (Id: 2017.1.00121.S) exist which was observed on 20 September 2018 with a 5.9215 AU geocentric distance. After analyzing the data using {\tt CASA}, we find the rotational absorption line of water at 183.310 GHz above 4$\sigma$ statistical significance. The absorption feature of the water line detected by TP data is coming from the atmosphere of Earth. The absorption spectrum is shown in Fig.~\ref{fig:water1}. The absorption line occurs due to strong artifacts in the T$_{sys}$ and target spectra coming from the LO of the Precipitable Water Vapour (PWV) radiometers. In Jovian atmosphere, the absorption line of water is found only in the troposphere of Jupiter \citep{Bj86} but it can not be transported from the Jovian tropospheres to the stratospheres due to a cold trap at the tropopause \citep{Be20}. We cannot believe the Jovian tropospheric water is detected in 183.310 GHz. The water vapour is mainly coming from the stratosphere of Jupiter with emission feature which we detect first time using ALMA at 183.310 GHz with ACA data. 

\subsection{Abundances of \ce{H2O} in Jovian atmosphere}
\label{sec:col}	
The column density is a statistically measurable quantity that is used to understand the chemical and physical nature of the object, as well as analysis of the chemical composition, isotopomer ratios, molecular abundance, kinetic temperature, and isomer ratio \citep{man15}. The total column density ($N_{tot}$) of the detected molecule can be calculated by,\\

$N_{tot} = \frac{3h}{8 \pi^3 |{\mathbf \mu_{lu}}|^2}
\frac{Q_{rot}}{g_u} \exp\left(\frac{E_u}{kT_{ex}}\right)
\times\left[\exp\left(\frac{h\nu}{kT_{ex}}\right) - 1\right]^{-1}
\int\tau_\nu dv.$ ---------(1)\\ 

where, $h$ = plank constant, $|{\mathbf \mu_{lu}}|^2$ = dipole matrix element (such that $|{\mathbf \mu_{jk}}|^2 \equiv S\mu^2$), Q$_{rot}$ = $\sum_i g_i \exp\left(-\frac{E_i}{kT}\right)$ = rotationl partiton function, $g_u$ = degeneracy level, $E_u$ = upper energy of detected species, $T_{ex} = \frac{h\nu / k}{\ln \left( \frac{n_l \, g_u}{n_u \, g_l} \right)}$ = excitation temperature, and $\tau_\nu$ = optical depth. The values of statistical parameters $E_{u}$, $g_{u}$, and $Q_{rot}$ are taken from CDMS\footnote{\href{https://cdms.astro.uni-koeln.de/cgi-bin/cdmssearch}{https://cdms.astro.uni-koeln.de/cgi-bin/cdmssearch}}. Equation.~1 can be used directly to calculate the total column density for emission lines where the integral of the optical depth over velocity can be derived from the spectrum. Using equation.~1, we calculate the column density corresponding to the \ce{H2O} emission line is N(H$_{2}$O)$\sim$4$\times$10$^{15}$ cm$^{-2}$ with $T_{ex}$ = 64.94 K. The column density of H$_{2}$O corresponds to the fractional abundance relative to \ce{H2} is f(\ce{H2O})$\sim$ 4$\times$10$^{-9}$. The calculated mixing ratio of water in the stratosphere of Jupiter using ALMA at 183.310 GHz is very close with \citet{Be20}.\\

\section{Discussion}
\label{sec:discussion} 
In this paper, we present the first spectroscopic detection of the rotational emission line of water at frequency $\nu$ = 183.310 GHz ($\geq$3$\sigma$ significance) with J = 3$_{1,3}$--2$_{2,2}$ molecular transitions. The statistical column density of detected water emission line is $\sim$ 4$\times$10$^{15}$ cm$^{-2}$, and column density of H$_{2}$O corresponds to the fractional abundance relative to \ce{H2} is f(\ce{H2O})$\sim$ 4$\times$10$^{-9}$. Our calculated column density of water 4$\times$10$^{15}$ cm$^{-2}$ is very close to the previously calculated column density of water in the stratosphere of Jupiter, which is (2.0$\pm$0.5)$\times$10$^{15}$ cm$^{-2}$ \citep{Lel02}. The close match of column density indicates that the detaction of water vapour reported in the present work is mainly coming from the stratosphere of Jupiter.

The detection of water on Jupiter fills a critical space to understand the molecular chemistry in the solar system. Earlier, water vapour was found in the atmosphere of solar planet Mars \citep{Fou11}, Venus \citep{Don93}, Jovian moon Europa \citep{Har11} and Saturn moon Titan \citep{Feu99}. Recently the water molecular emission line was also detected from the thin atmosphere of Moon at 6$\mu$m using SOFIA \citep{Hon20}. The detection of an emission line of water in stratosphere of Jupiter may be critical in the formation of some complex bio compounds as well as the formation of life. The formation and evolution of life with carbon-based metabolism are difficult without the presence of water. The presence of HCN, CO, \ce{H2O}, and other carbon-based molecules gives confidence to the possibility of the formation of amino acids in the stratosphere of Jupiter according to Miller and Urey's experiment \citep{Mil53}. More spectroscopic studies of the complex molecular transition lines in the stratosphere of Jupiter will help to confirm the nature and formation mechanism of the detected trace gases.

\section*{Acknowledgement}
This paper makes use of the following ALMA data: ADS /JAO.ALMA\#2017.1.00121.S. ALMA is a partnership of ESO (representing its member states), NSF (USA), and NINS (Japan), together with NRC (Canada), MOST and ASIAA (Taiwan), and KASI (Republic of Korea), in cooperation with the Republic of Chile. The Joint ALMA Observatory is operated by ESO, AUI/NRAO, and NAOJ.

\section*{Data Availability Statement}
The data that support the plots within this paper and other findings of this study are available from the corresponding author upon reasonable request. The raw ALMA data are publicly available at \href{https://almascience.nao.ac.jp/asax/}{https://almascience.nao.ac.jp/asax/} (project id : 2017.1.00121.S).


\begin{thebibliography}{99}
	
	\bibitem[\protect\citeauthoryear{Benmahi et al.}{2020}]{Be20} Bilal, B., et al. 2020, A\&A, A140, 9.
	
	\bibitem[\protect\citeauthoryear{Bjoraker et al.}{1986}]{Bj86} Bjoraker, G.L., Larson H.p., \& Kunde, V.G., 1986, ApJ, 311, 1058.
	
	\bibitem[\protect\citeauthoryear{Bjoraker}{2020}]{Bjo20}Bjoraker G. L., 2020, Nature Astronomy, 4, 558. 
	
	\bibitem[\protect\citeauthoryear{Butler}{2012}]{But12}Butler B., 2012, ALMA Memo 594, ALMA Memo Series, NRAO.
	
	\bibitem[\protect\citeauthoryear{Donahue \& Hodges}{1993}]{Don93} Donahue, T., \& Hodges, R. R. 1993, Geophysical Research Letters. 20. 
	
	
	\bibitem[\protect\citeauthoryear{Encrenaz et al.}{1995}]{enc95}Encrenaz, T., Lellouch, E., Cernicharo, J., Paubert, G., \& Gulkis, S., 1995, Icarus, 113, 110.
	
	\bibitem[\protect\citeauthoryear{Feuchtgruber et al.}{1999}]{Feu99}Feuchtgruber H., et al. 1999, ESA SP-427, 133.
	
	\bibitem[\protect\citeauthoryear{Fletcher et al.}{2009}]{Fle09} Fletcher, L.N., Orton, G.S., Teanby, N.A., \& Irwin, P.G.J., 2009, Icarus, 202.
	
	
	\bibitem[\protect\citeauthoryear{Fouchet et al.}{2011}]{Fou11}Fouchet T., Moreno R., Lellouch E., Formisano V., Giuranna M., \& Montmessin F., 2011, Planetary and Space Science, 59, 683.
	
	\bibitem[\protect\citeauthoryear{Friedson et al.}{2002}]{Fri02}Friedson A. J., Wong A.-S., \& Yung Y. L., 2002, Geophysical Research Letters, 30. 
	
	\bibitem[\protect\citeauthoryear{Gautier et al.}{1981}]{Gau81}Gautier D., Conrath B., Flasar M., Hanel R., Kunde V., Chedin A., \& Scott N., 1981, Journal of Geophysical Research, 86, 8713.
	
	
	\bibitem[\protect\citeauthoryear{Hartogh et al.}{2011}]{Har11} Hartogh, P., et al. 2011, A\& A, 532.
	
	\bibitem[\protect\citeauthoryear{Honniball et al.}{2020}]{Hon20}Honniball C., Lucey P., Li S., Shenoy S., Hibbitts C., Hurley D., \& Farrell W., 2020, Nature Astronomy, 1.
	
	\bibitem[\protect\citeauthoryear{Hueso \& S{\'a}nchez-Lavega}{2002}]{Hue06} Hueso R., \& S{\'a}nchez-Lavega A., 2006, Nature, 442, 428.
	
	\bibitem[\protect\citeauthoryear{Kunde et al.}{2004}]{Kun04}Kunde V. G. et al. 2004, Science, 305, 1582.
	
	\bibitem[\protect\citeauthoryear{Lellouch et al.}{2002}]{Lel02}Lellouch E. et al. 2002, Icarus, 159, 112.	
	
	\bibitem[\protect\citeauthoryear{Lellouch et al.}{1995}]{Lel95}Lellouch E. et al. 1995,
	Nature, 373, 592.
	
	\bibitem[\protect\citeauthoryear{Loeffler \& Hudson}{2017}]{Loe17}Loeffler M. J., \& Hudson, R. j., 2017, Icarus, 302, 418.
	
	\bibitem[\protect\citeauthoryear{Li et al.}{2020}]{Li202}Li C. et al. 2020,  Nature Astronomy, 4, 609. 
	
	\bibitem[\protect\citeauthoryear{Mangum \& Shirley}{2015}]{man15}Mangum, J., \& Shirley, Y., 2015, Publications of the Astronomical Society of the Pacific, 127. 
	
	\bibitem[\protect\citeauthoryear{Mason}{1980}]{Mas80}Mason B. J., 1980, Contemporary Physics, 21, 381.
	
	\bibitem[\protect\citeauthoryear{Miller}{1953}]{Mil53} Miller S. L., 1953, Science, 117, 5.
	
	\bibitem[\protect\citeauthoryear{Moses et al.}{2004}]{mos04} Moses, J.~I. et al. 2004, The Planet, Satellites and Magnetosphere, 129.
	
	\bibitem[\protect\citeauthoryear{Moreno et al.}{2003}]{Mor03}Moreno R., Marten, A., Matthews H. E., \& Biraud Y., 2003, Planetary and Space Science, 51, 591.
	
	\bibitem[\protect\citeauthoryear{Nol et al.}{1995}]{Nol95}Noll K. S. et al., 1995, Science, 267, 1307.		
	
\end{thebibliography}
\end{document}